\documentclass[%
preprint, 
 amsmath,amssymb,
 prfluid
]{revtex4-2}

\usepackage{graphicx}
\usepackage{dcolumn}
\usepackage{bm}

\usepackage{amsmath,rotating}
\usepackage{dashrule}
\usepackage{caption}
\usepackage{subcaption}
\usepackage{multirow}
\usepackage{array}
\usepackage{makecell}

\usepackage{todonotes}
\usepackage{soul}
\usepackage{gensymb} 

\usepackage{booktabs}   
\usepackage{multirow}   
\usepackage{array}      

\begin{document}

\preprint{}

\title{\textbf{Large eddy simulations of conjugate heat transfer in boundary layers over laser-scanned ice roughness} 
}%

\author{F. Zabaleta}
\affiliation{Center for Turbulence Research, Stanford University}

\author{B. Bornhoft}
\affiliation{Cadence Design Systems\,}

\author{S. S. Jain}
\affiliation{Flow Physics and Computational Science Lab, GWW School of Mechanical Engineering, Georgia Institute of Technology}

\author{S. T. Bose}
\affiliation{Cadence Design Systems}
\affiliation{Institute for Computational and Mathematical Engineering (ICME), Stanford University}

\author{P. Moin}
\affiliation{Center for Turbulence Research, Stanford University\,}

\date{\today}

\begin{abstract}
Accurate modeling of ice accretion is important for the safe and efficient design of aircraft and wind turbine systems. Heat transfer predictions obtained from the fluid flow solvers are used as input in ice accretion codes. In glaze ice conditions, the freezing rates and resulting ice shapes are highly sensitive to the input values of the heat transfer coefficient. Hence, an accurate prediction of heat transfer on iced airfoils is crucial for correctly predicting the ice accretion process \citep{liuExperimentalInvestigationUnsteady2018,hansonImprovedCFDApproach2017}. In this study, we perform conjugate heat transfer (CHT) simulations using wall-modeled large-eddy simulation (WMLES) over surfaces characterized by ice roughness. The results show that WMLES with CHT accurately captures surface temperature distributions and heat fluxes across a range of roughness geometries. For the cases considered, large roughness-to-boundary-layer thickness ratios disrupt outer-layer similarity, leading to substantial errors in estimating equivalent sandgrain roughness when applying traditional empirical models based on surface statistics. The simulations further show that local heat fluxes vary significantly across roughness elements due to the low thermal conductivity of the solid; in particular, roughness crests exhibit reduced fluxes in contrast to slopes and valleys. Notably, as roughness height increases, the wall heat flux at the crest diminishes, even leading to heat flux reversal in some cases, where thermal energy is transferred from the fluid to the solid. These effects are not captured in isothermal wall simulations, which overestimate the Stanton number, especially at roughness peaks. By enabling the calculation of Stanton number using heat flux distributions, not directly available in the experiments, the present simulations augment experimental results and highlight the importance of including solid conduction effects for accurately modeling heat transfer over rough, low-conductivity surfaces such as ice. 
\end{abstract}

\maketitle

\section{Introduction}
The impingement of supercooled water droplets on aircraft wings and other lifting surfaces can lead to ice formation, altering their aerodynamic properties. Aircraft icing results in a significant reduction of lift and a rapid increase in drag, contributing to several aviation accidents \citep{valarezoMaximumLiftDegradation1993}. Beyond affecting aerodynamic performance, icing can compromise propulsion efficiency by accumulating on components such as nacelles, fan blades, and inlet ducts. Atmospheric ice crystals can adhere to engine interiors, potentially causing total power loss and damage \citep{masonIceParticleThreat2006}. Ice accretion also impacts wind turbine performance, where ice-induced roughness reduces power generation \citep{gaoWindTurbineIcing2021}. Therefore, accurate modeling of ice accretion is essential to ensure the safety and efficiency of new aircraft designs.

Two primary types of ice, rime and glaze ice, accrete on aerodynamic surfaces. Rime ice typically occurs at low flight temperatures ($<-20^{\circ} C$), low speeds, low atmospheric liquid water concentration, and small droplet sizes \cite{gentAircraftIcing2000}. In rime ice conditions, cooling mechanisms remove all the latent heat of solidification from the impinged supercooled water droplets, freezing them completely upon impact. Under these conditions, the ice shape is determined primarily by the impingement distribution of the droplets and is almost independent of the heat transfer process. In contrast, glaze ice typically forms at near-freezing temperatures, higher speeds, elevated liquid water concentration, and larger droplet sizes. In glaze conditions, not all droplets freeze on impact; a portion of them freeze and the remaining liquid runs back along the airfoil, either freezing further downstream or departing from the surface. This type of ice often appears translucent and can form structures known as horns, which severely degrade aerodynamic performance. In this regime, the rate of ice accretion is highly sensitive to the heat-transfer capacity needed to remove the latent heat of solidification from the droplets.

Numerical ice accretion modeling is typically performed in the following steps: First, a steady-state Reynolds-averaged Navier-Stokes (RANS) solver is used to compute the surface flow field. Next, either an Eulerian or Lagrangian droplet transport model is employed to compute the rate of impact of droplets on the body surface \citep{ozcerAnsysBombardier1st2022a,wrightGlennICECapabilitiesResults2022a,zabaletaLargeEddySimulationSupercooled2024}. Information obtained from these models is used as an input to solve a continuity and an energy conservation equation for a water film over the surface, calculating the liquid flow and the freezing rate \citep{messingerEquilibriumTemperatureUnheated1953,aliagaThirdgenerationInflightIcing2007}. Next, the outer mold line of the iced surface is updated, and the domain is remeshed. This process of computing the flow field, droplet collection, and calculating the growth of ice is typically referred to as one shot. When this process is repeated multiple times, it is known as a multishot calculation. In general, ice accretion codes based on a RANS framework are unable to capture certain stochastic features observed in real ice. To address these limitations, several recent models introduce stochasticity directly into the ice accretion framework in an effort to better predict local roughness features~\cite{bellostaMultiStepStochasticApproach2024,ozcerMultiShotIcingSimulations2019}.

In glaze conditions, the freezing rates and resulting ice shapes are highly sensitive to the values of the heat transfer coefficient ($h$) obtained from the simulations and used as input to the ice accretion model \citep{kontogiannisSensitivityGlazeIce2018,hansonImprovedCFDApproach2017}. The convective heat transfer coefficient is often obtained from running isothermal wall simulations and then calculating $h$ using $q_w=h \times (T_{w}-T_{aw})$, where $q_w$ is the wall heat flux, $T_w$ is the wall temperature, and $T_{aw}$ is the adiabatic wall temperature \citep{wrightGlennICECapabilitiesResults2022a}.
The effects of surface roughness on the convective heat‐transfer coefficient \(h\) are typically incorporated into models via one of three approaches: (1) by specifying an equivalent sand‐grain roughness height \(k_s\) to modify the turbulence closure in the RANS equations \citep{ozcerFENSAPICENumericalPrediction2011,cacciaNumericalSimulationsIce2023}; (2) by applying empirical correlations developed for icing conditions \citep{mcclainModelIceAccretion2021}; and (3) by employing the discrete‐element roughness method to perform a subgrid representation of the roughness elements \citep{hansonComputationalInvestigationConvective2017a}.
Comparisons with experiments have shown that ice accretion codes can overpredict heat transfer coefficients by almost 400\% \citep{hanTransientHeatTransfer2014}. Results from the Ice Prediction Workshop I revealed a significant disparity in heat transfer coefficients among participants, with peaks ranging from 500 W/m$^2$K to 2500 W/m$^2$K \citep{laurendeauSummary1stAIAA2022}. For these reasons
, accurate prediction of heat transfer on iced airfoils is crucial for correctly predicting the ice accretion process \citep{liuExperimentalInvestigationUnsteady2018}. 

Roughness influences iced surfaces through multiple mechanisms. In the early stages of ice accretion, surface roughness can trip the boundary layer to transition prematurely, increasing both skin friction and heat transfer. As roughness develops, it plays a significant role in enhancing drag and convective heat transfer \citep{bellostaAssessingRelevantRoughness2024}, which affect droplet runback and the dissipation of latent heat. These enhancements are traditionally represented as vertical shifts relative to the logarithmic velocity and temperature profiles of smooth walls \citep{chungPredictingDragRough2021}.  


Roughness-resolved simulations of iced surfaces have been conducted, but these have primarily focused on aerodynamic degradation caused by to ice shape, rather than on heat transfer characteristics \citep{konigLatticeBoltzmannAnalysisThreeDimensional2015,bornhoftLargeeddySimulationsNACA230122024}. In this article, we present an alternative approach to simulating heat transfer in icing environments through a series of roughness-resolved numerical simulations. We employ a conjugate heat transfer (CHT) solver coupled with wall-modeled large-eddy simulation (WMLES) to model heat transfer in a developing rough-wall boundary layer. The numerical setup replicates the experiments conducted by Ref. \onlinecite{mccarrellConvectionSurfacesIce2018}, who performed heat transfer measurements over a series of flat plates featuring ice-characterized roughness. Heat was generated by Mylar plates installed below the testing plate, leading to temperature variations on the rough surfaces that were captured using an infrared (IR) camera installed above it. Here, we compare the temperature distributions and wall heat fluxes obtained from the coupled CHT/WMLES simulations with the experimental measurements. One of the primary goals of this study is to evaluate the ability of WMLES to predict heat transfer on iced surfaces, examining the influence of the wall model in the representation of the temperature boundary layer and the grid resolution required for accurate heat transfer predictions. This study also assesses the role of solid thermal conductivity in modulating heat transfer over rough surfaces and explores its implications for icing simulations. In parallel, we evaluate the performance and limitations of traditional roughness representations based on an equivalent sandgrain roughness. Finally, the simulations presented here serve as a validation case for the CHT/WMLES framework and provide a reference dataset to support the refinement of existing RANS-based models used in icing prediction.

The article is structured as follows. Section \ref{sec:numModel} details the CHT/WMLES model, including the governing equations, wall models, and the treatment of different timescales between solid and fluid domains. Section \ref{sec:expData} described the experimental data used to validate the numerical model and Section \ref{sec:compSetup} describes the computational setup. Validation of the numerical model is provided in Section \ref{sec:Validation}. Section \ref{sec:roughnessChars} characterizes the influence of surface roughness on the flow and assesses the accuracy of empirical models used for predicting $k_s$ for the cases considered. Section \ref{sec:heatTransfer} discusses the heat fluxes obtained from the simulation and the impact of solid conductivity on the Stanton number ($St$) distribution. Finally, Section \ref{sec:conclusions} summarizes the study and presents the conclusions. 

\section{Numerical model} \label{sec:numModel}

\subsection{Considerations for heat transfer modeling}
Before describing the specific numerical approach, we first assess the necessity of using conjugate heat transfer for the present simulations. In order to do so, the dimensionless Biot number ($Bi$) is considered. The Biot number compares the internal thermal resistance due to conduction within a solid to the external resistance due to convection at its surface:
\begin{equation}
    Bi = \frac{h L}{\lambda_s},
\end{equation}
where $\lambda_s$ is the thermal conductivity of the solid, and $L$ is a characteristic length scale of the solid.

For $Bi \ll 0.1$, the internal conduction resistance is much smaller than the convective resistance at the surface, resulting in negligible temperature gradients within the solid. In this regime, the solid can be assumed isothermal and detailed CHT modeling is generally not necessary. However, for $Bi \gtrsim 0.1$, temperature gradients within the solid become significant, since the conduction resistance is comparable to or exceeds the convective resistance. In such cases, CHT is required to resolve the temperature distribution within the solid and accurately predict the surface temperature and heat flux.

In the present work, we estimate the Biot number based on the experimental conditions described in Section~\ref{sec:expData}, neglecting the effects of radiation (a conservative assumption). From the experiments, $h \approx 50~\text{W}/(\text{m}^2{\cdot}\text{K})$, $\lambda_s = 0.162~\text{W}/(\text{m}{\cdot}\text{K})$, and the representative length scale is taken as the maximum peak-to-trough roughness height, $k_z \approx 0.01~\text{m}$. For this conditions, $Bi \approx 3$, indicating substantial temperature gradients within the roughness elements. Therefore, CHT modeling is essential to accurately simulate the heat transfer process over these rough surfaces.

\subsection{Governing equations}
The computational code employed in this study is charLES \citep{bresLargeeddySimulationsCoannular2018}, a low-dissipation, second-order accurate in space, and fifth-order accurate in time,  finite-volume solver. In this work, the low-pass filtered, compressible Navier-Stokes equations for mass, momentum, and total energy,
\begin{equation}
    \frac{\partial \overline{\rho}}{\partial t} + \frac{\partial \overline{\rho}\tilde{u_i}}{\partial x_i} = 0, 
\end{equation}
\begin{equation}
    \frac{\partial \overline{\rho}\tilde{u_i}}{\partial t} + \frac{\partial \overline{\rho}\tilde{u_i}\tilde{u_i}}{\partial x_i} = - \frac{\partial \overline{p}}{\partial x_i} + \frac{\partial \tilde{\tau}_{ij}}{\partial x_i} - \frac{\partial \tilde{\tau}^{sgs}_{ij}}{\partial x_i},
\end{equation}
\begin{equation}
    \frac{\partial \overline{E}}{\partial t} + \frac{\partial \tilde{u_i} \overline{E}}{\partial x_i} + \frac{\partial \tilde{u_i}\overline{p}}  {\partial x_i} = \frac{\partial \tilde{\tau}_{ij}\tilde{u_i}}{\partial x_i} - \frac{\partial \tilde{\tau}^{sgs}_{ij}\tilde{u_i}}{\partial x_i} + \frac{\partial}{\partial x_i}\left( \lambda \frac{\partial \overline{T}}{\partial x_i}\right) - \frac{\partial \tilde{Q}^{sgs}_{i}}{\partial x_i},
\end{equation}
are solved, where the operators $\overline{(\cdot)}$ and $\tilde{(\cdot)}$ represent filtered and Favre (density-weighted) filtered quantities, respectively. In these equations, $u_i$ stands for the velocity vector, $p$ represents the pressure, $\rho$ is the density, and $\tau_{ij}$ and $\tau^{sgs}_{ij}$ represent the viscous and subgrid stresses, respectively. The subgrid terms are closed using the dynamic Smagorinsky model \citep{moinDynamicSubgridScale1991}, for which eulerian time averaging is used to determine the Smagorinsky coefficient. The stress tensors are calculated as,
\begin{equation}
    \tau_{ij} = 2\mu \left( S_{ij}- \frac{1}{3}\delta_{ij}S_{kk} \right)
\end{equation}
\begin{equation}
    \tau_{ij}^{sgs} = -2 C_s \rho \Delta^2 |S| \left(S_{ij} - \frac{1}{3}S_{kk}\delta_{ij}\right) + \frac{1}{3}\tau_{kk}^{sgs}\delta_{ij},
\end{equation}
with
\begin{equation}
    |S| = \sqrt{2 S_{mn} S_{mn}}, \qquad
S_{ij} = \frac{1}{2} \left(\frac{\partial \tilde{u}_i}{\partial x_j} + \frac{\partial \tilde{u}_j}{\partial x_i}\right), \quad \tau_{kk}^{sgs} \approx 0 .
\end{equation}
$E$ stands for the total energy, $T$ is the temperature, and $Q_i^{sgs}$ represents the subgrid heat flux. A constant molecular Prandtl number ($Pr=0.7$) is used to compute the thermal conductivity ($\lambda$), while the subgrid heat flux is computed using a constant turbulent Prandtl number approximation ($Pr_t=0.9$), as follows,
\begin{equation}
    Q_i^{sgs} = \frac{1}{Pr_t} C_s \rho \Delta^2 |S| \frac{\partial \overline{T}}{\partial x_i}.
\end{equation}
Pressure, density, and temperature are related by the ideal gas law. Viscosity and thermal conductivity are modeled as functions of temperature using a power law with a coefficient of 0.75. Skew-symmetric operators are employed to conserve kinetic energy in the inviscid, zero Mach number limit, and the scheme also approximately preserves entropy in the inviscid, adiabatic limit \citep{honeinHigherEntropyConservation2004,chandrashekarKineticEnergyPreserving2013}. This discretization has been shown to be suitable for coarsely resolved LES of turbulent flows that are especially sensitive to numerical dissipation. 

On the solid domain, the unsteady heat conduction equation is solved to obtain the temperature distribution,
\begin{equation}
    \frac{\partial \rho_s c_{p,s} T}{\partial t} + \frac{\partial}{\partial x_j}\left( \lambda_s \frac{\partial T}{\partial x_j}\right) = 0,
\end{equation}
and the heat flux at the fluid-solid interface is imposed based on the LES calculations,
\begin{equation}
    \lambda_s \left.\frac{\partial T}{\partial n} \right|_w = q_{w,\text{LES}} + q_{\text{radiation}},
\end{equation}
where $\rho_s$ is the solid density, $c_{p,s}$ is the solid heat capacity, $T$ represents the solid temperature, $\lambda_s$ denotes the thermal conductivity of the solid, $q_{w,\text{LES}}$ stands for the wall heat flux imposed from the LES calculations, and $q_{\text{radiation}} = \sigma \epsilon (T_w^4 - T_\infty^4)$ is the radiative heat flux, where $\sigma$ is the Stefan–Boltzmann constant, $\epsilon=0.95$ is the emissivity of the ABS plastic, and $T_\infty$ is the free-stream temperature. Although convection is the dominant mode of heat transfer for the cases presented in this article, the radiative heat flux accounts for approximately \(18\%\) of the total heat flux. Neglecting radiation leads to an overprediction of the surface temperature. The solution procedures in the fluid and solid domain are tightly coupled; i.e., temperatures and heat-fluxes at the interface are updated every time-step based on the results from the solid and fluid domains, respectively.

\subsection{Wall modeling}
Due to the high Reynolds number in many engineering applications, resolving the viscous sublayer using LES becomes prohibitively expensive. To obtain accurate results with reasonable computational resources, wall models are used to represent the shear stress and the heat flux at the solid wall \citep{boseWallModeledLargeEddySimulation2018}. In the laminar region, the shear stress is imposed based on Blasius boundary layer equations \citep{gonzalezWallstressModelingLaminar2020} as
\begin{equation}
    \tau_w = 0.332\sqrt{\frac{\rho \mu U_\infty^3}{x}},
\end{equation}
where $U_\infty$ is the free-stream velocity, and the heat flux is imposed  assuming equilibrium in the wall-normal direction. For the turbulent region, an equilibrium wall model is employed. We assume that the pressure gradient balances the advective terms, which leads to the constant stress layer approximation,
\begin{equation}
    \frac{\partial \tau}{\partial n} = \frac{\partial }{\partial n}\left( (\mu+\mu_{t\text{,wm}}) \frac{\partial \tilde{u}_t}{\partial n} \right) = 0,
    \label{eq:WM}
\end{equation}
where $\tilde{u}_t$ and $\tilde{u}_n$ are the velocity components tangential and normal to the wall, respectively, and $\mu_{t\text{,wm}}$ is the wall-model eddy viscosity. In this study, we use an algebraic approximation of the wall model derived from the integration of  Eq. (\ref{eq:WM}). The heat flux at the wall is calculated assuming equilibrium in the wall-normal direction: 
\begin{equation}
    \frac{d}{d n}\left( (\mu+\mu_{t\text{,wm}}) \tilde{u}_t \frac{d \tilde{u}_t}{d n} + \left( \frac{\mu}{Pr}+\frac{\mu_{t\text{,wm}}}{Pr_t}\right)  \frac{d \overline{T}}{d n}   \right)=0.
\end{equation}
For more detail on the wall model, the reader is referred to \citep{lehmkuhlLargeeddySimulationPractical2018}. On the interface boundaries between fluid and solid domains, the temperature is imposed based on the solution of the heat equation obtained from the solid side.

\subsection{Treatment of solid thermal transients}

There are substantial differences in the timescales for heat transfer in the solid and the fluid. While the characteristic thermal diffusion time within the solid is given by $t_s = \rho_s c_{p,s} L^2 / \lambda_s \approx 1300~\mathrm{s}$, the characteristic time associated with turbulent transport in the boundary layer at the end of the domain is $t_\text{fluid} = \delta_{99}/u_\tau \approx 0.1~\mathrm{s}$. This large disparity indicates that temperature variations in the solid occur much more slowly than in the fluid. Consequently, it is necessary to employ an approach to accelerate the thermal soaking of the solid. In this study, we employ the equalized timescale method \cite{diefenthalThermomechanicalAnalysisTransient2017}, which accelerates thermal soaking by reducing the thermal inertia of the solid. By modifying only the thermal inertia, the Biot number remains unchanged, ensuring that the quasi-steady temperature distribution is unaffected \citep{cuiInvestigationEffectConjugate2022}. Thermal inertia is initially reduced by a fixed factor, $\beta$. The simulations are started with a thermal inertia set to $\beta \rho c_p = 10^{-4} \rho c_p$. Once a statistically stationary temperature state is reached in the solid, $\beta$ is gradually increased to eliminate unphysical temperature fluctuations caused by the initially low thermal inertia. This gradual increase continues until $\beta$ reaches 1. At $\beta = 1$, the surface temperature becomes nearly insensitive to flow fluctuations. A description of the temperature evolution and the effect of $\beta$ is presented in Appendix \ref{Ap:betaEffect}.

\begin{figure}
    \centering
    \includegraphics[width=\linewidth]{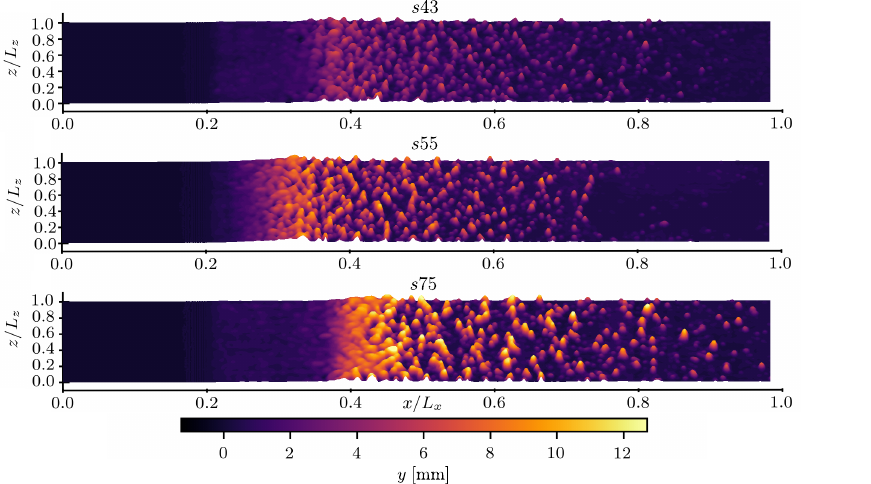}
    \caption{Surface elevation maps of the $s43$, $s55$ and $s75$ geometries.}
    \label{fig:geometries}
\end{figure}

\begin{figure}
    \centering
    \includegraphics[width=\linewidth]{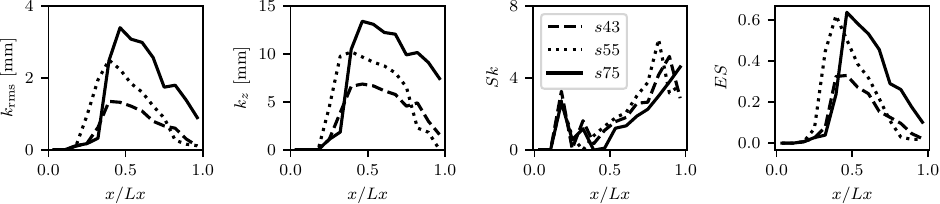}
    \caption{Root-mean-square roughness height ($k_{rms}$), 2N-point-average ($N=5$) peak-to-trough roughness height ($k_z$), skewness ($Sk$), and effective slope ($ES$) as a function of the streamwise coordinate for the three geometries.}
    \label{fig:surfaces_stats}
\end{figure}

\section{Experimental data}\label{sec:expData}
The experimental data used to validate the model correspond to the experiments described in Ref.~\onlinecite{mccarrellConvectionSurfacesIce2018}, which were conducted in the Baylor University Subsonic Wind Tunnel. In this study, the geometries labeled 113012.05, 112912.02, and 113012.04 will be considered. The geometries correspond to the unwrapped suction section of iced NACA 0012 airfoils created in the Icing Research Tunnel at NASA Glenn Research Center. They result from exposing the airfoils at a freestream temperature of $-2.37\degree $C to supercooled large droplets (SLD) conditions for 43, 55, and 75 seconds, respectively. The iced geometries were laser scanned and unwrapped using the methodology of Ref.  \onlinecite{mcclainAssessmentIceShape2013}. The three unwrapped geometries are depicted in Figure \ref{fig:geometries} and colored by surface elevation. Throughout this article, $x$, $y$, and $z$ coordinates represent to the streamwise, wall-normal, and spanwise directions, respectively. Hereafter, the geometries will be referred by their exposure time as $s43$, $s55$, and $s75$. 

\begin{figure}
    \centering
    \includegraphics[width=0.6\textwidth]{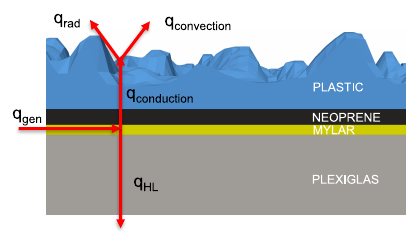}
    \caption{Schematic of the experimental setup, showing the four layers present in the experiment and the heat fluxes (adapted from Ref. \onlinecite{mccarrellConvectionSurfacesIce2018})}
    \label{fig:Exp_setup}
\end{figure}

The rough surfaces were 3D printed in ABS plastic, placed over a neoprene layer and mounted over a Plexiglass support. In between the Plexiglass and the neoprene, five Mylar plates provided heating to the surface, with an approximate heat flux of $q_{\text{gen}} \approx$ 500 W/m$^2$. Thermocouples were installed at the Mylar plate, at the bottom of the plexiglas layer, and in the free-stream flow. They were used to estimate the heating lost through the Plexiglass base ($q_{\text{HL}}$). A schematic of the experimental setup, showing the four layers present in the experiment and the heat fluxes, is presented in Figure \ref{fig:Exp_setup}. In the heat transfer experiments, the geometry was scaled by a factor of 10, and the testing velocity was reduced proportionally to maintain the Reynolds number achieved during the accretion process. The three plates are \(L_x = 37.5\)~inches long and \(L_z = 8\)~inches wide. The free-stream velocity was set to \(U_\infty = 6.67\)~m/s. The Reynolds and Mach numbers of the experiment are $Re_{Lx} \approx 4.2\times10^5$ and $Ma \approx 0.02$. The boundary layer is not tripped, resulting in a laminar boundary layer that undergoes transition caused by the beginning of the rough region. The transition location ($X_t$), which coincides approximately with the beginning of roughness, obtained from skin friction measurements, is $X_t/L_x = 0.4$, $0.3$, and $0.41$ for the $s43$, $s55$, and $s75$ geometries, respectively.

Each surface initially exhibits a smooth region followed by the onset of roughness at different streamwise locations. These differences arise from the stochastic nature of ice formation, as each surface represents a distinct realization under identical experimental conditions. Figure~\ref{fig:surfaces_stats} shows the streamwise evolution of spanwise averaged root-mean-square roughness height ($k_\text{rms}$), the spanwise averaged 2$N$-point-average ($N=5$) peak-to-trough roughness height ($k_z$), spanwise averaged skewness ($Sk$), and spanwise averaged effective slope ($ES$) for the three geometries. For definitions of these parameters, the reader is referred to Ref.~\onlinecite{chungPredictingDragRough2021}. Small frequency variations of these parameters, resulting from the limited width of the domain, are removed by applying a filter of length $L_x/20$. The roughness height reaches its maximum near the beginning of the rough region and progressively decreases downstream. The maximum roughness height is located at $X_{k,\max}/L_x = 0.44, 0.37$ and 0.5 for $s43$, $s55$, and $s75$ geometries, respectively, with peaks of $k_\text{rms}$ and $k_z$ coinciding. In $s43$ and $s55$, the surface transitions back to an almost flat plate towards the downstream end. In contrast, roughness is still present at the end of $s75$. For all geometries, the faster decrease of $k_\text{rms}$ compared to $k_z$, along with the associated increase in skewness, indicates that roughness elements become more scattered towards the downstream region. Such spatial variability in roughness height distribution is characteristic of ice accretion processes \citep{sotomayor-zakharovStatisticalAnalysisSurface2024}.

We have selected this dataset to validate our model because it provides high-resolution measurements of temperature distributions at various accretion times. This enables us to evaluate the model at different stages and to assess the effects of changing roughness heights. However, it is important to acknowledge certain differences between these experiments and actual icing conditions. First, in real icing, heat is released at the surface primarily due to the latent heat of solidification, whereas in the experiments considered here, heat is supplied from below the surface. This may affect the role of conduction on the heat-transfer process. Second, the experiments are conducted at a free-stream temperature of $293.15~\mathrm{K}$ ($20^\circ$C), while icing typically occurs at much lower ambient temperatures. Finally, the thermal conductivity of ABS plastic used in the experiments is nearly an order of magnitude lower than that of ice. Although both are low-conductivity materials, this difference can lead to varying temperature gradients and alter the significance of conductive heat transfer within the solid.

Despite these limitations, we consider this dataset to be highly valuable for studying heat transfer over surfaces with ice-like roughness. It enables us to validate convective heat transfer calculations under different accretion stages and to investigate the role of the solid's conduction on the convective heat transfer for low-conductivity materials.


\section{Computational setup}\label{sec:compSetup}

\begin{figure}
    \centering
    \includegraphics[width=\textwidth]{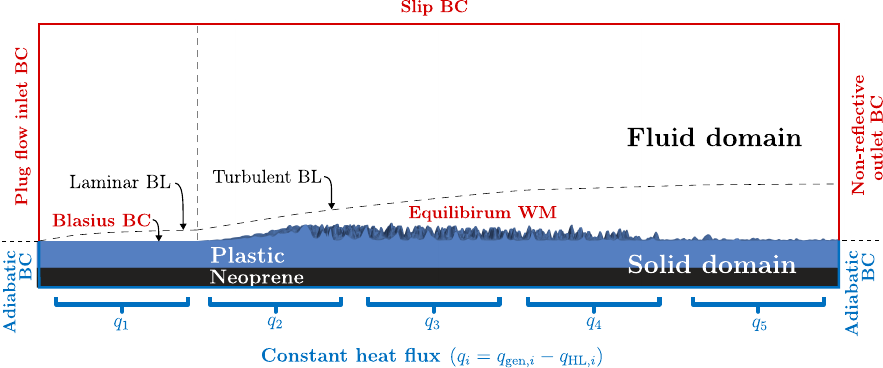}
    \caption{Schematic of the computational domain. Red text indicates the fluid domain boundary conditions, and blue text indicates the solid domain boundary conditions. Vertical dimensions are not to scale.}
    \label{fig:comp_domain}
\end{figure}

The computational setup replicates the experimental setup described on the previous section. A schematic of the computational domain, including boundary conditions, is presented in Figure \ref{fig:comp_domain}. The fluid domain includes the totality of the plate, while the solid domain includes the plastic and neoprene layers. On the fluid domain, the inlet is set with the mean free-stream velocity, and for the outlet a non-reflecting characteristic boundary condition is used \citep{poinsotBoundaryConditionsDirect1992}. Slip boundary conditions are used at the top and the side walls. At the plate, wall models are used to calculate the shear stress. For $x \leq X_t$ a Blasius boundary condition is employed, while for $x > X_t$\ an algebraic equilibrium wall model is used. The temperature at the surface results from the solution of the heat equation in the solid domain. The free-stream temperature is set based on the experiments to $T_\infty = 293.15$ K. The solid domain considers only the neoprene and plastic layers. The material properties for neoprene are specified as $\rho_n = 1230$ kg/m$^3$, $c_{p,n} = 2200$ J/(kg·K), and $\lambda_{s,n} = 0.175$ W/(m·K). For ABS plastic, the properties are set to $\rho_p = 1070$ kg/m$^3$, $c_{p,p} = 1990$ J/(kg·K), and $\lambda_{s,p} = 0.162$ W/(m·K). The ratio of solid to fluid conductivity is $\lambda_{s,n}/\lambda = 7$ and $\lambda_{s,p}/\lambda = 6.5$. A sensitivity analysis involving a 20\% variation in the material properties of neoprene and ABS plastic was conducted, showing only minor variations in surface temperatures (within 5\% of the experimental uncertainty). At the bottom boundary, constant heat fluxes for each heating plate are set based on heat generated in the plates minus the heat lost through the Plexiglass base $(q_{\text{gen}} - q_{\text{HL}})$. The heating on each individual panel is calculated based on the reported current and voltage 
on the individual heaters as $q_{\text{gen}} = EV/A_h$, where $E$ is the current through heater, $V$ is the voltage across heater, and $A_h$ is the area of the heater. Lateral walls are set using adiabatic boundary conditions. At the interface with the fluid domain, heat fluxes are imposed based on the solution of the fluid domain and the radiative fluxes.

\begin{table}[htb]
\begin{ruledtabular}
\centering
\begin{tabular}{@{} l l c c c c c @{}}
\multirow{2}{*}{Geometry} 
  & \multirow{2}{*}{Grid} 
    & \multicolumn{1}{c}{\(N_{\mathrm{cv}}\) [Mcv]} 
    & \multicolumn{1}{c}{\(k_{\mathrm{rms}}/\Delta_{\min}\)} 
    & \multicolumn{1}{c}{\(k_{z}/\Delta_{\min}\)} 
    & \multicolumn{1}{c}{\(\delta_{99}/\Delta_{\min}\)}
    & \multicolumn{1}{c}{\(y^+\)}  \\

  & 
    & \multicolumn{1}{c}{[fluid, solid]} 
    & \multicolumn{1}{c}{[max, mean]} 
    & \multicolumn{1}{c}{[max, mean]} 
    & \multicolumn{1}{c}{[\((0.1,\,0.4,\,0.7)\,L_x\)]}
    & \multicolumn{1}{c}{[max, mean]}  \\
\midrule

\multirow{3}{*}{s43} 
  & Coarse &  1.5, 0.2  &  1.1, 0.6  &   5.6, 3.7   &  1.9, 8, 17   & 68, 20    \\
  & Medium &  6.3, 2.0  &  2.2, 1.1  &  11, 7.4     &  3.9, 15, 34   & 34, 10   \\
  & Fine   & 25, 15     &  4.5, 2.3  &  22, 15      &  7.7, 30, 67   & 17, 5   \\
\midrule

\multirow{3}{*}{s55} 
  & Coarse &  1.6, 0.2  &  1.9, 0.9  &   7.9, 5.7   &  1.9, 13, 22  &   68, 20  \\
  & Medium &  6.6, 1.9  &  3.7, 1.8  &  16, 11      &  3.9, 27, 44  &   34, 10  \\
  & Fine   & 27, 16     &  7.5, 3.6  &  32, 23      &  7.7, 54, 88  &   17, 5  \\
\midrule

\multirow{3}{*}{s75} 
  & Coarse &  1.7, 0.2  &  2.6, 1.3  &  11, 7.3     &  1.9, 10, 24   & 70, 22   \\
  & Medium &  6.9, 2.2  &  5.2, 2.6  &  21, 15      &  3.9, 19, 49   & 35, 11  \\
  & Fine   & 28, 17     & 11, 5.2    &  42, 29      &  7.7, 38, 97   & 17, 6   \\
\end{tabular}
\caption{Grid refinement details: cell counts \(N_{\mathrm{cv}}\) in the fluid and solid domain; ratio of mean and maximum root‐mean‐square roughness height (\(k_{\mathrm{rms}}\)) and two‐point‐average peak‐to‐trough roughness height (\(k_{z}\)) to minimum grid length scale (\(\Delta_{\min}\)); and ratio of boundary layer height (\(\delta_{99}\)) to \(\Delta_{\min}\) at \(x/L_x=0.1,\,0.4,\,0.7\).}
\label{tab:meshes}
\end{ruledtabular}
\end{table}

\begin{figure}
    \centering
    \includegraphics[width=0.7\textwidth]{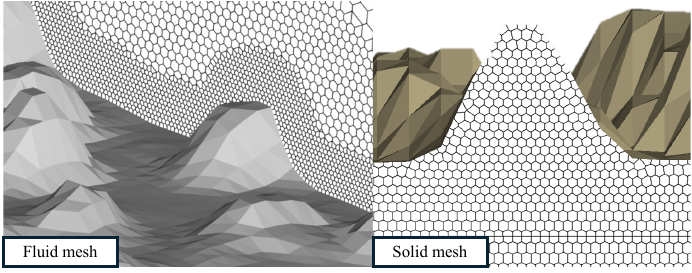}    
    \captionsetup{singlelinecheck=false}
    \caption{Computational mesh over two slices in roughness region, in the fluid domain (left) and the solid domain (center), for the fine mesh.}
    \label{fig:Comp_mesh}
\end{figure}

The body-fitted computational mesh based on Voronoi diagrams is constructed using the Lloyd algorithm \citep{duConvergenceLloydAlgorithm2006}. Three meshes of sequential refinements levels are used in the simulations. The details of the meshes for the three geometries are presented in Table \ref{tab:meshes}. Each refinement level is achieved via halving the characteristic cell size. The meshes are refined at the surface and progressively coarsened away from it by reducing one refinement level every 15 layers. In addition to the wall-normal clustering, using isotropic grid cells ensures adequate resolution of the roughness elements in the streamwise and spanwise directions. The fine mesh resulted in an average of $k_{\text{rms}}/\Delta_{\min} = 2.3-5.2$ grid elements per $k_{\text{rms}}$ and 15--29 elements per $k_z$, where $\Delta_{\min}$ represent the minimum cell size (located at the wall). Figure \ref{fig:Comp_mesh} presents images of the fine mesh over two planes, highlighting the resolution near the rough surface (for the fluid mesh) and inside a roughness element (for the solid mesh). These resolutions allow us to capture the flow around roughness elements as well as temperature variations within them. As shown in Table \ref{tab:meshes}, the resolution within the laminar region of the boundary layer ($\delta_{99}/\Delta_{\text{min}}$ at $0.1L_x$) is insufficient to accurately resolve the thin thermal laminar boundary layer. 
However, as demonstrated in Section \ref{sec:Validation}, this limitation affects only a small portion of the domain and does not compromise the accuracy of the downstream results. Increasing the resolution in this region would significantly increase the computational cost without notable benefit.

\section{Validation}\label{sec:Validation}
\begin{figure}
    \centering
    \includegraphics[width=\textwidth]{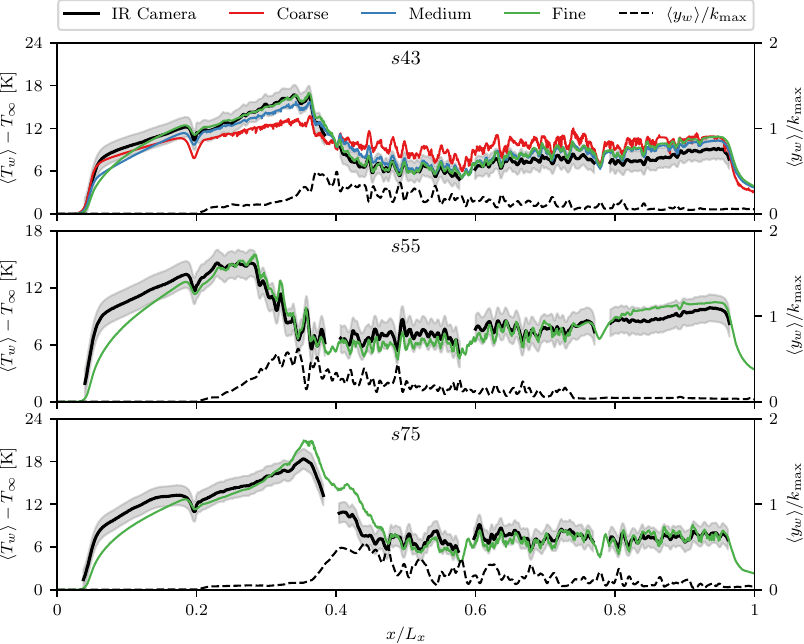}
    \caption{Spanwise-averaged surface temperature profiles over a band of thickness $\Delta z/L_z = 1/2$ located at the plate centerline obtained with the coarse, medium, and fine meshes, as well as with the IR camera for geometry $s43$ and with fine mesh, for $s55$ and $s75$. $\langle . \rangle$ operator indicates spanwise-averaging. The gray are indicates the uncertainty of the measurements with the IR camera ($\pm1.4$ K). Dashed line represents spanwise averaged surface elevation.}
    \label{fig:T_profiles}
\end{figure}

To validate the model, temperature distributions over the plate surface are compared with the experimental data. Surface temperature data is used to validate the model because it is the only variable related to heat transfer directly measured in the experiments. Other quantities, such as heat transfer coefficient or $St$, were derived from the temperature measurements. Figure \ref{fig:T_profiles} presents spanwise-averaged surface temperature profiles over a central band of thickness $\Delta z/L_z = 1/2$ for the three geometries, comparing numerical predictions with IR camera measurements. Spanwise averaging is restricted to the central half of the plate to minimize the influence of boundary effects observed in the experiments. To maintain clarity, results for the three mesh resolutions are shown only for geometry $s43$, which exhibits the smallest roughness height, and results for the finest mesh are shown for $s55$ and $s75$. Adequate resolution for $s43$ case is expected to ensure sufficient resolution for the other geometries, which feature larger roughness elements. Results from the medium and fine meshes for $s43$ are closely aligned, indicating mesh convergence in surface temperature predictions. For the finer meshes, the computed temperatures generally fall within the experimental uncertainty ($\pm 1.4$ K) across most of the domain. Local maxima and minima in the temperature profile are accurately captured in the roughness region, highlighting the model’s effectiveness in predicting heat transfer over rough surfaces. 

All three geometries exhibit discrepancies with the experimental data in the laminar region ($x/L_x \leq 0.2$), likely due to a combination of insufficient resolution in the early portion of the boundary layer and the presence of small pressure gradients induced by the leading edge of the mounting structure, which is not included in the numerical model. 
Since the primary focus of this study is on heat transfer in the rough region, these differences are considered acceptable. Resolving the thin laminar boundary layer in detail would require a significantly finer mesh than that employed here, with substantial computational cost increase \citep{Slotnick2014}. 
Moreover, the observed discrepancies in the upstream laminar region do not affect the results in the downstream region of interest. 
In addition, case $s75$ presents some discrepancies with the experimental data near $x/L_x=0.4$, where the numerical model overpredicts the wall temperature peak at the onset of the roughness region. These differences diminish quickly within the initial portion of the rough surface. These results show that a mesh resolution of approximately $k_\text{rms}/\Delta_{\min} \approx 2$ is enough to capture heat transfer and temperature variations over ice roughness. These results are in line with resolution requirements to predict aerodynamic loads on iced airfoils \citep{bornhoftLargeeddySimulationsNACA230122024}. 
Overall, the model shows strong agreement with experimental observations, indicating its high accuracy in representing heat transfer features.

\begin{figure}
    \centering
    \includegraphics[width=\textwidth]{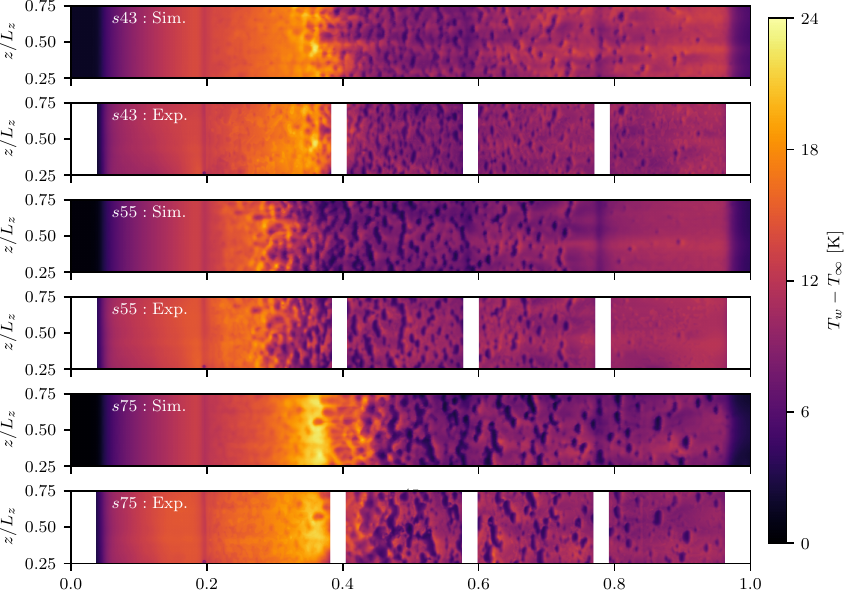}
    \caption{Surface temperature distribution for $s43$, $s55$ and $s75$ geometries obtained with the numerical model and from the experiments with the IR camera.} 
    \label{fig:T_maps}
\end{figure}

Figure \ref{fig:T_maps} presents the surface temperature distribution for the three geometries obtained with the numerical model with the fine mesh and from the experiments with the IR camera in the center portion of the plate ($0.25 \leq z/L_z \leq 0.75$). Qualitatively,  temperature distributions from the numerical model and from the IR camera are very similar. Temperature increases almost monotonically in the smooth portion of the plate (the small dip at $x/L_x=0.2$ results from an unheated area between heating plates), reaching maximum temperatures where roughness starts. The presence of roughness induces transition, which significantly increases the heat fluxes and rapidly reduces the temperature of the plate. In the roughness region, the temperature profile stabilizes, displaying large temperature gradients due to the presence of roughness elements. Temperature variations are larger on $s75$ and $s55$, where roughness elements are larger. Cooler patches correspond to the tops of these roughness elements, while warmer regions form within the troughs between them. The peaks of the roughness elements experience direct exposure to colder airflow and are farther from the heat source, resulting in lower temperatures. Conversely, the troughs are both closer to the heat source and shielded from direct airflow, contributing to higher temperatures. Additionally, the area immediately downstream of the roughness elements exhibits further increased surface temperatures due to the sheltering effect created by these protrusions. 
As the roughness ends, the temperature distribution becomes mostly uniform. 

\begin{figure}
    \centering
    \includegraphics{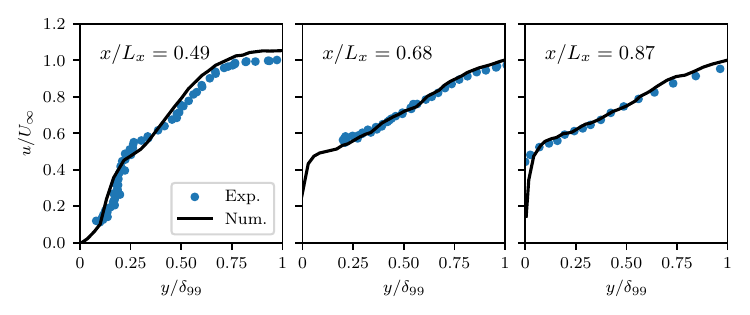}
    \caption{Velocity profiles at $x/L_x=0.29,\;0.49,$ and 0.68 (left, center, right, respectively) for $s55$ geometry obtained with the numerical model and from the experiments.}
    \label{fig:velocitys55}
\end{figure}

To further validate the model, Figure \ref{fig:velocitys55} compares measured and simulated velocity profiles in the rough region for geometry $s55$. The numerical results exhibit excellent agreement with the experimental data. Similar accuracy was also achieved for the remaining geometries (not shown herein), confirming the robustness of the numerical model across different roughness conditions.


\section{Roughness characterizaton}\label{sec:roughnessChars}
Most icing simulations, as well as experiments involving artificial ice shapes, are typically performed using smooth geometries with superimposed sand-grain-type roughness \citep{galliaAutomaticRoughnessCharacterization2023,ozoroskiDetailsAdheringTesting2025}. Thus, it is valuable to analyze and characterize the roughness features of these geometries and estimate the equivalent sand-grain roughness height under the present flow conditions. It is important to emphasize that $k_s$ is a hydrodynamic length scale and does not necessarily correlate directly with the geometrical height of the roughness elements \citep{flackImportantParametersPredictive2022}. Mean velocity profiles over rough walls can be expressed as a function of $k_s$ as:
\begin{equation}
    U^+ = \frac{1}{\kappa}\ln\left(\frac{y^+}{k_s^+}\right) + B 
    \label{eq:LoWks}
\end{equation}
where $B$ depends on the value of $k_s^+$. In the fully rough regime, $B$ approaches a constant value that is a function of the roughness geometry. $\kappa$ is the von K\'arm\'an constant, and the superscript ‘+’ denotes variables normalized using inner (viscous) scaling. The wall-normal distance is corrected by the zero-plane displacement, which is estimated as the center of drag on the wall \citep{jacksonDisplacementHeightLogarithmic1981}. Alternatively, velocity profiles can be expressed using the roughness function $\Delta U^+$, which quantifies the downward shift of the log law relative to a smooth wall at matched wall-normal distance $y^+$:
\begin{equation}
    U^+ =  \frac{1}{\kappa}\ln\left(y^+\right) + C - \Delta U^+
    \label{eq:LoWDeltaU}
\end{equation}
where $C$ is the log-law intercept for smooth walls. These two parameters are equivalent and can be related by:
\begin{equation}
    \Delta U^+ =\frac{1}{\kappa} \ln(k^+_s) + C - B.
    \label{eq:DeltaU-ks}
\end{equation}

The existence of a logarithmic layer requires that $y > y_k$ and $y \ll \delta$ (or $y_k\ll\delta$), where $y_k$ is the height of the roughness sublayer, defined as the region where the flow is directly influenced by the local roughness topology, and is typically estimated as $y_k \approx 3k_s$ \citep{flackExaminationCriticalRoughness2007}. The existence of outer-layer similarity has been historically associated with small roughness-to-boundary-layer-height ratios. \citet{jimenezTurbulentFlowsRough2004} suggested that outer-layer similarity holds for $k/\delta < 1/40$, although subsequent studies have demonstrated its validity for $k/\delta < 1/19$ \citep{flackExaminationCriticalRoughness2007}, and even up to $k/\delta \approx 1/7$ \citep{chungPredictingDragRough2021}.

\begin{figure}
    \centering
    \includegraphics{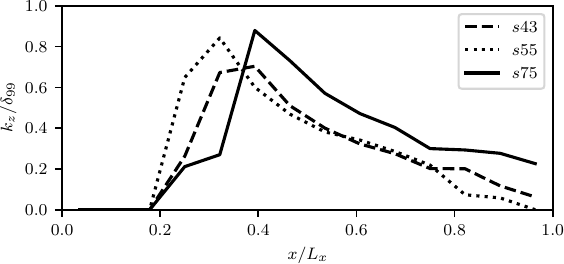}
    \caption{$k_z/\delta_{99}$ ratio as a function of the streamwise coordinate for cases $s43$, $s55$ and $s75$}
    \label{fig:kz_over_delta}
\end{figure}

Figure \ref{fig:kz_over_delta} shows the variation of $k_z/\delta_{99}$ along the plate for the three geometries. Values of $k_z/\delta_{99}$ reach approximately $0.6-0.8$ in regions where the roughness height is largest, and then decrease downstream as the boundary layer thickens. These values far exceed the thresholds typically associated with the existence of outer-layer similarity or even a logarithmic layer. Additional velocity deficit profiles (presented in Appendix \ref{Ap:velProfiles}) confirm that outer-layer similarity is not recovered within the length of the domain. The high $k_z/\delta_{99}$ ratios indicate that the flow over these surfaces is more representative of an obstacle regime rather than a canonical rough-wall regime. In this regime, the influence of roughness elements extends across the entire boundary layer, and parameters such as $\Delta U^+$ and $k_s^+$ no longer provide a meaningful description of the velocity profiles. 

\begin{figure}
    \centering
    \includegraphics[width=\linewidth]{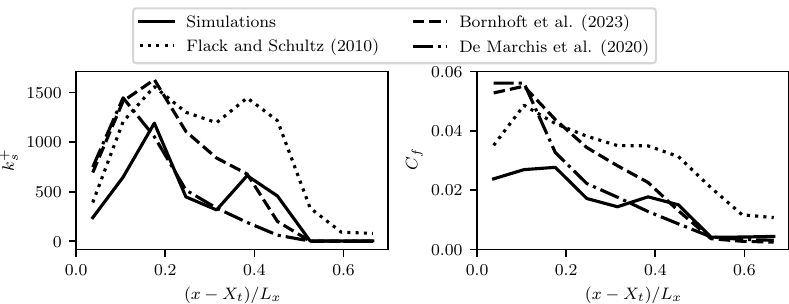}
    \caption{Left: Computed values of $k_s^+$ for case $s55$, obtained from the simulations using Equation~(\ref{eq:ks_friction}) and predicted using the correlations from Equations~(\ref{eq:ks_FlackSchultz}), (\ref{eq:ks_Bornhoft}), and (\ref{eq:ks_DeMarchis}). Right: $C_f$ calculated directly from the simulations and estimated indirectly via the $k_s^+$ values obtained from Equations~(\ref{eq:ks_FlackSchultz}), (\ref{eq:ks_Bornhoft}), and (\ref{eq:ks_DeMarchis}).} 
    \label{fig:ks-computed}
\end{figure}

As noted at the beginning of this section, $k_s$ is widely employed in icing simulations, despite the fact that $k/\delta$ often exceeds the established limits for the existence of outer-layer similarity. For this reason, even though this parameter may not be physically meaningful in the present context, we still compute an equivalent sand-grain roughness, that is, a value of $k_s$ that would yield the same wall shear stress as the actual rough surface. Since a logarithmic layer is absent in many of the velocity profiles, Equation~(\ref{eq:LoWks}) cannot be used to estimate $k_s^+$ directly. However, Equation~(\ref{eq:LoWDeltaU}) can be rearranged to express the roughness function $\Delta U^+$ in terms of the difference between friction coefficients for smooth and rough walls at matched $\delta^+$:
\begin{equation}
\Delta U^{+} = \sqrt{\frac{2}{C_{f,S}}} - \sqrt{\frac{2}{C_{f,R}}},
\label{eq:ks_friction}
\end{equation}
where $C_{f,S}$ and $C_{f,R}$ denote the skin friction coefficients for the smooth and rough walls, respectively. This value of $\Delta U^+$ can then be used in Equation~(\ref{eq:DeltaU-ks}) to obtain the equivalent roughness height $k_s^+$. Figure \ref{fig:ks-computed} (left) presents, for geometry $s55$, the computed values of $k_s^+$ obtained using Equation~(\ref{eq:ks_friction}) and predicted using the models from Flack and Schultz \cite{flackReviewHydraulicRoughness2010},
\begin{equation}
    k_s^+ = 4.43k_{\text{rms}}^+(1+Sk)^{1.37};
    \label{eq:ks_FlackSchultz}
\end{equation}
from Bornhoft et al. \cite{bornhoftVelocityTransformationRoughwallbounded2023},
\begin{equation}
\Delta U^{+}=c_1 \ln \left(c_2 k_{r m s}^{+} ES\right)\left[\left(c_3-1\right) \tanh \left(c_4 s_k\right)+1\right] e^{-c_5 ES^{c_6}}
\label{eq:ks_Bornhoft}
\end{equation}
where $c_1-c_6$ are empirical coefficients; and from De Marchis et al. \cite{demarchisLargeEddySimulations2020},
\begin{equation}
\Delta U^{+}=\frac{1}{\kappa} \log\left(k_\textrm{rms}^+ ES\right) + 3.5.
\label{eq:ks_DeMarchis}
\end{equation}

To compute $\Delta U^{+}$ using these models, spanwise-averaged roughness statistics are employed as inputs. A filter with a length of $L_x/20$ is applied to remove high-frequency oscillations in roughness statistics resulting from the limited domain span. All three models tend to overpredict the values of $k_s^+$, which is expected given that they were originally calibrated for significantly lower values of $k/\delta$ than those present in the current simulations. The models proposed by Bornhoft et al. \cite{bornhoftVelocityTransformationRoughwallbounded2023} and De Marchis et al. \cite{demarchisLargeEddySimulations2020} capture the overall trend of the computed $k_s^+$ values reasonably well. In particular, they reproduce the sharp rise at the beginning of the rough region, the location of the peak at $(x-X_t)/L_x = 0.2$, and the gradual decay toward negligible values by $(x-X_t)/L_x = 0.53$. However, both models fail to predict the secondary peak observed at $(x-X_t)/L_x = 0.4$. In contrast, the model from Flack and Schultz \cite{flackReviewHydraulicRoughness2010} captures this secondary peak but substantially overestimates $k_s^+$ for $(x-X_t)/L_x > 0.2$. Figure~\ref{fig:ks-computed} (right) shows the values of $C_f$ obtained directly from simulations (accounting for both skin friction and form drag) alongside those estimated indirectly from the $k_s^+$ values predicted by the previously discussed models. As expected, the overestimation of $k_s^+$ leads directly to an overestimation of $C_f$ on the rough portion of the plate. The largest discrepancy was observed using Equation~(\ref{eq:ks_FlackSchultz}), with an average error of approximately 138\%. The models by Bornhoft et al. \cite{bornhoftVelocityTransformationRoughwallbounded2023} and De Marchis et al. \cite{demarchisLargeEddySimulations2020} performed comparatively better, yielding mean estimation errors in $C_f$ of 61\% and 43\%, respectively. However, in the final portion of the plate, where the ratio \(k_z/\delta_{99}\) is smaller, these two models yield lower mean errors of 26\% and 23\%, respectively. These results highlight that under the present conditions, predicting surface roughness via available $k_s^+$ models can result in significant inaccuracies in estimating both $k_s^+$ and consequently wall-shear stress. It is important to emphasize that this limitation is not inherent to the models themselves; rather, these models were specifically developed for rough-wall conditions where outer-layer similarity holds. Furthermore, models to estimate $k_s^+$ were developed for fully turbulent regimes, not for transitional regimes such as those encountered in the initial portion of the rough region. In icing scenarios, roughness heights can grow rapidly and reach magnitudes comparable to the boundary layer thickness, thus violating the outer-layer similarity hypothesis and significantly reducing the accuracy of conventional roughness parametrizations. Hence the approach of using smooth surfaces superimposed with sand-grain roughness may not be very appropriate for simulating icing conditions.

\section{Heat transfer}\label{sec:heatTransfer}

\begin{figure}
    \centering
    \includegraphics[width=\textwidth]{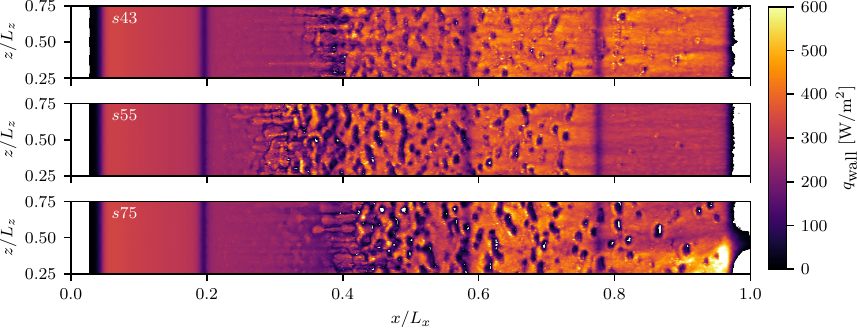}
    \caption{Heat-flux distribution obtained with the numerical model for $s43$, $s55$ and $s75$ geometries (top, center and bottom, respectively). White areas indicate negative values of $q_w$.}
    \vspace{-.5cm}
    \label{fig:q-comp}
\end{figure}

One of the primary goals of this work is to characterize convective heat transfer over surfaces with ice roughness. To this end, heat-flux distributions obtained from the numerical model for the three geometries are presented in Figure \ref{fig:q-comp}. The heat flux shows a clear transition from the smooth to the rough region of the plate. In the smooth section, the distribution is relatively uniform and corresponds to the imposed heat flux from the heating plates, reduced by radiative losses. The vertical dark bands in this region (and throughout the figure) reflect the spacing between adjacent heating panels. As the flow encounters the roughness, elongated streak-like structures appear, oriented predominantly in the streamwise direction. These features represent the thermal imprint of individual roughness elements and their wakes, and suggest localized enhancements in convective heat transfer due to flow disturbances and recirculation induced by the protrusions. The persistence and spacing of these streaks highlight the role of roughness-induced coherent structures in promoting mixing and disrupting the uniformity of the thermal boundary layer.

In the rough region, the heat flux becomes highly heterogeneous. The highest values are typically observed on the windward sides of the roughness elements, where colder fluid impinges creating high-temperature gradients. The leeward sides exhibit slightly lower heat fluxes. The tops of the roughness elements often show the lowest heat fluxes, even reaching negative values  on the leeward side of the crest (heat is transferred from the fluid to the solid). As the roughness height increases (from case $s43$ to $s75$), the wall heat flux at the roughness peaks decreases, and a larger number of peaks exhibit negative heat fluxes (white areas in Figure \ref{fig:q-comp}). These results contrast with those from studies of rough isothermal walls, where the highest heat fluxes typically occur at the top of the roughness elements due to their greater exposure to colder ambient air \citep{macdonaldRoughnessEffectsTurbulent2019a}. When thermal conduction within the solid is accounted for, however, this trend reverses. Although the top of the elements are indeed exposed to colder fluid, they also become colder themselves due to limited thermal conduction from the base. This effect has been observed in previous studies \cite{orlandiDNSConjugateHeat2016,orlandiTransitionalNaturalConvection2017}; however, those studies did not report heat flux reversal at the roughness crests, likely because the roughness height considered was smaller. As our results show, increasing roughness height and accounting for low solid conduction can lead to stronger variation of heat fluxes, and even a local inversion of the heat flux direction.

\begin{figure}
    \centering
    \includegraphics[width=\textwidth]{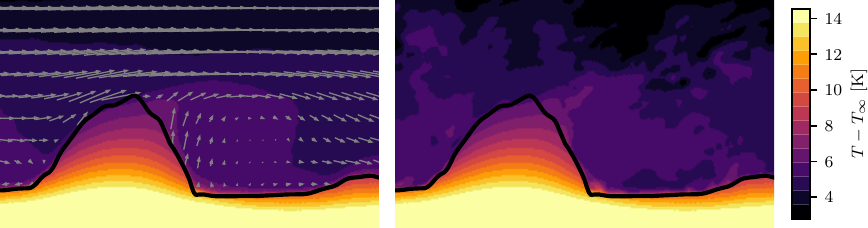}
    \caption{Mean (left) and instantaneous (right) temperature fields on the fluid and solid domain, superimposed with mean velocity vectors around a roughness element. Black line represents solid-fluid interface.  }
    \label{fig:avgTandT}
\end{figure}

Figure \ref{fig:avgTandT} presents the mean and instantaneous temperature fields in the solid and fluid regions of the domain around a roughness element, with mean velocity vectors superimposed. On the solid side, temperature decreases with height, with the windward face exhibiting lower temperatures than the leeward face due to enhanced cooling from the direct impact of the cooler fluid flow. Behind the roughness element, a recirculation zone forms, which entrains hot air from the base (which is in contact with the hottest parts of the surface) and transports it upward along the leeward side. This process leads to locally elevated temperatures on the leeward face. In some roughness elements, these local temperatures exceed the corresponding surface temperature, resulting in a reversal of the heat flux direction at the wall.

\begin{figure}
    \centering
    \includegraphics{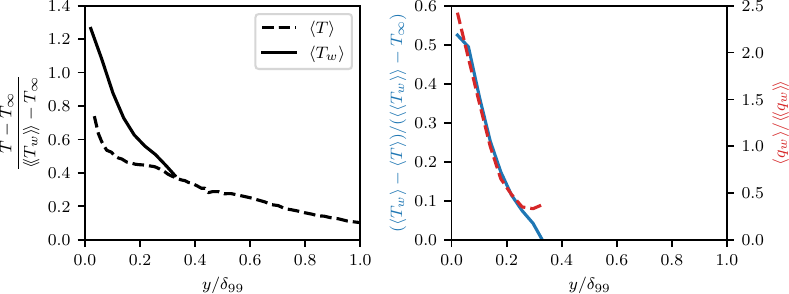}
    \caption{Left: Vertical profiles of the spanwise-averaged fluid temperature, $\langle T \rangle$, and the averaged wall temperature, $\langle T_w \rangle$, at $x/L_x=0.6$ for geometry $s75$. Right: Vertical distribution of the temperature difference between the fluid and the wall, along with the corresponding normalized wall heat flux. In this plot, $\langle\! \langle {\cdot} \rangle\! \rangle$ indicates the quantity averaged successively over the spanwise and wall-normal coordinates.}
    \label{fig:temp_profiles}
\end{figure}

To quantitatively characterize this effect, we compare spanwise-averaged profiles of wall temperature, $\langle T_w \rangle$, with spanwise-averaged profiles of fluid temperature, $\langle T \rangle$. It is important to note that, because we are considering rough surfaces, $\langle T_w \rangle = \langle T_w \rangle(x,y)$ varies with both the streamwise and wall-normal coordinates. Figure~\ref{fig:temp_profiles} (left) presents the vertical profiles of $\langle T \rangle$ and $\langle T_w \rangle$ at $x/L_x = 0.6$ for geometry $s75$. The fluid temperature decreases gradually with wall-normal distance, whereas the wall temperature drops more rapidly. This indicates that, for low-conductivity materials, heat is transported more efficiently by convection in the fluid than by conduction within the solid. The reduction in the local temperature difference between the wall and the adjacent fluid with increasing wall-normal distance, shown in Figure~\ref{fig:temp_profiles} (right), closely correlates with the decrease in wall heat flux, showing that $\langle q_w \rangle \sim \langle T_w \rangle - \langle T \rangle$. This reduction in temperature difference underscores the critical role of internal conduction within roughness elements in controlling convective heat transfer over low-conductivity rough surfaces. Capturing these coupled mechanisms requires conjugate heat transfer simulations that resolve both solid and fluid thermal transport. The results presented in Figure~\ref{fig:temp_profiles} for $s75$ at $x/L_x = 0.6$ are representative of the temperature profiles at other locations in the plate where roughness is present. Similar behaviours can be observed for $s43$ and $s55$. In these geometries, because roughness elements are shorter, the reduction in temperature difference is not as significant as for $s75$. This is reflected on the heat-flux maps where these two geometries have smaller regions with negative values of $q_w$.

\subsection{Stanton number distribution}
Convective heat transfer is described by the Stanton number, which measures the ratio of heat transferred into a fluid to the thermal capacity of the fluid. The $St$ is defined as
\begin{equation}
    St = \frac{h}{\rho c_p U_\infty} = \frac{q_{w}}{\Delta T \rho c_p U_\infty},
\end{equation}
where $\Delta T = T_{w}\, -\, T_\infty$ is the difference between the surface temperature and the free-stream temperature. Measuring heat fluxes over rough surfaces presents significant challenges; therefore, to compute maps of $St$ (or $h$), the experiments assumed that $q_{w} = q_{\text{gen}} - q_{\text{HL}} - q_{\text{radiation}}$, where $q_{\text{gen}}$, $q_{\text{HL}}$ and $q_{\text{radiation}}$ are the heat generated by the heating plate, the heat lost through the Plexiglass plate, and the heat lost due to radiation, respectively. We refer to this heat flux $q_0$, which has uniform values over each heating plate, and are equal to zero in between them. The Stanton number based on $q_0$ is denoted as $St_0$, which can be determined using $\Delta T$ from either the experiments ($St_{0,\textrm{exp}}$) or the simulations ($St_{0,\textrm{num}}$). However, this assumption for $q_w$ is not required in simulations, as the actual values of $q_w$ are directly accessible. Consequently, $St$ can be calculated using these local $q_w$ values without assuming uniform heat flux across the heating plates, and we designate the Stanton number calculated this way as $St_{\textrm{num}}$.

\begin{figure}
    \centering
    \includegraphics[width=\textwidth]{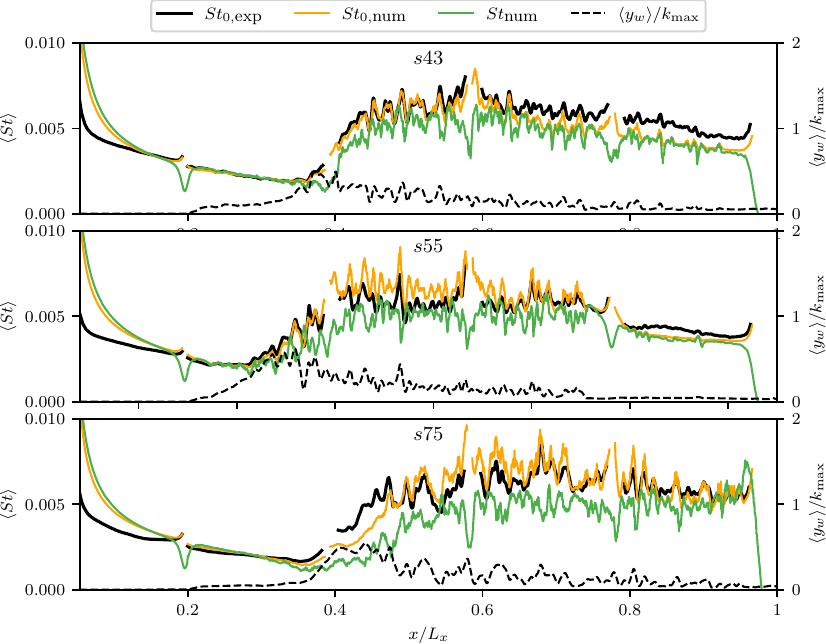}
    \caption{Spanwise-averaged $St$ profiles over a central band of thickness $\Delta z/L_z = 1/2$ for the three geometries, obtained with the numerical model using a uniform heat flux distribution $q_0$ ($St_{0,\textrm{num}}$), as well as using the heat fluxes obtained from the simulation ($St_{\textrm{num}}$), superimposed with the $St$ obtained using the experimental temperatures and $q_0$ ($St_{0,\textrm{exp}}$). Discontinuities in black and orange lines correspond to the gaps between heating plates, where values of $q_w$ and $T_w$ were not reported in the experiments. Dashed line represents spanwise averaged surface elevation.}
    \vspace{-.5cm}
    \label{fig:St_profiles}
\end{figure}

Figure \ref{fig:St_profiles} shows the spanwise-averaged profiles of $St_{0,\textrm{exp}}$, $St_{0,\textrm{num}}$, and $St_{\textrm{num}}$, computed over a central band with thickness $\Delta z/L_z = 1/2$. The $St_{0,\textrm{num}}$ profile closely matches the experimental data. This is expected given the strong agreement in surface temperature distributions and the fact that both profiles use the same values of $q_w$ in the calculation of $St$. The $St$ profile calculated with the heat fluxes from the simulations ($St_{\textrm{num}}$) matches the other two profiles in the smooth regions ($x/L_x \leq 0.3$ and $x/L_x \geq 0.7$), but significantly differs in the rough region. In the smooth region, distributions of $q_w$ and $T_w$ are relatively uniform, and therefore $St_{0,\textrm{num}} \approx St_{\textrm{num}}$. In the rough region, $St_{0,\textrm{exp}}$ and $St_{0,\textrm{num}}$ display local maxima at the top of the roughness elements; however, in $St_{\textrm{num}}$ maxima shift towards the front area of the roughness elements. As shown in Figure \ref{fig:q-comp}, heat fluxes present significant variability in the rough region. The highest values of $q_w$ over the rough region are located in windward slopes, while on the peaks heat fluxes are significantly reduced, even reaching negative values. For $St$ distribution calculated using $q_0$ (uniform over each heating plate), peaks of $St$ coincide with peaks in roughness height, where $\Delta T$ is small. In the calculation of $St$ with $q_w$ from the simulations, peaks in $St$ are shifted toward the front region of the roughness elements, where the ratio of $q_w/\Delta T$ is larger. 

These simulations show that low solid conductivity materials (such as plastic or ice) alter heat transfer from surfaces with roughness. Important variations in surface temperature within roughness elements alter temperature gradients near the surface, affecting heat fluxes and Stanton number distributions. This effect is particularly relevant in these experiments, where the low thermal conductivity of ABS plastic restricts heat conduction to the peaks of the roughness elements, which in combination with them being more exposed to the colder free-stream flow, leads to a significantly lower temperature relative to their base. 
On the smooth regions, this effect is absent, and the influence of low-conductivity materials on convective heat transfer remains negligible.

\subsection{Comparison with isothermal walls}
Finally, we compare the $St$ values obtained with the CHT model to those obtained using isothermal walls. For icing applications, heat transfer is usually estimated using isothermal walls, ignoring conduction effects on the ice shape. The assumption of isothermal walls is valid for materials of infinite conductivity, and it breaks down for low-conductivity materials due to interaction of the flow with regions of the surface with large temperature gradients. The isothermal wall simulations are performed using the same mesh and setup as the CHT simulations, but using uniform $T_w$, equal to the mean temperature obtained from the CHT model. The initial unheated portion of the plate is represented on the isothermal simulation using an adiabatic wall. 

\begin{figure}
    \centering
    \includegraphics[width=\textwidth]{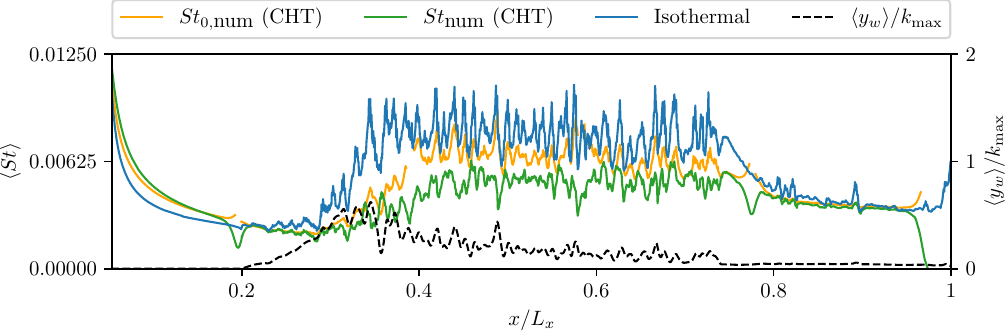}
    \caption{Spanwise-averaged $St$ profiles for $s55$ geometry over a central band of thickness $\Delta z/L_z = 1/2$, obtained with the numerical model using CHT ($St_{0,\textrm{num}}$ and $St_{\textrm{num}}$) and using isothermal walls.}
    \label{fig:isothermal-comp}
\end{figure}

The $St$ profiles for $s55$ geometry, obtained with the CHT model (both $St_{0,\textrm{num}}$ and $St_{\textrm{num}}$) and with isothermal walls are presented in Figure \ref{fig:isothermal-comp}. In contrast to $St_{\textrm{num}}$, the simulation with isothermal walls significantly overpredicts $St$ in the rough region. High-conductivity materials are known to yield higher $St$ than their low-conductivity counterparts \citep{orlandiDNSConjugateHeat2016,mccarrellConvectionSurfacesIce2018}. The simulation with isothermal wall resembles the $St_{0,\textrm{num}}$ profile, but with overall higher values of $St$. For the isothermal walls, peaks of $St$ coincide with peaks in roughness height, where maximum values of $q_w$ are located. These locations protrude deeper into the colder air flow, enhancing heat transfer. In the CHT simulations, as discussed in the previous subsection, peaks in $St$ are shifted toward the front area of the roughness elements. Limited conduction within the solid reduces the heat fluxes at the top of the roughness elements, leading to significantly different $St$ profiles in contrast to the isothermal simulation. Similar trends are observed for the $s43$ and $s75$ geometries, though they are omitted here for brevity.

\section{Conclusions}\label{sec:conclusions}

This study demonstrates that CHT combined with WMLES can accurately capture heat transfer over rough surfaces, provided that the grid resolution is sufficient to resolve the flow around the roughness elements. Across all three geometries considered, the simulations reproduced experimental temperature distributions accurately, except near the leading portion of the plate where resolution was insufficient to resolve the laminar boundary layer.

The examined cases exhibit large roughness-to-boundary-layer thickness ratios, ranging from $k_z/\delta_{99} \approx 0.8$ where roughness is the highest ($X_{k,\max}$) to $k_z/\delta_{99} \approx 0.2$ towards the end of the plate. These values indicate that the roughness effects penetrate the full boundary layer, disrupting outer-layer similarity and undermining the validity of empirical models based solely on surface statistics. As a result, such models tend to overpredict the equivalent sandgrain roughness ($k_s^+$) and wall-shear stress.

The simulations reveal that regions of high heat flux are concentrated on the windward faces of the roughness elements, while the leeward sides show moderately lower values. The crests of the roughness elements exhibit the lowest heat fluxes due to reduced temperature differences between the solid surface and the surrounding fluid. With increasing roughness height, the wall heat flux at the crest diminishes, even leading to heat flux reversal, where thermal energy is transferred from the fluid to the solid. This distribution contrasts with that observed in simulations using isothermal walls, highlighting the effect of thermal conduction within the solid. For low-conductivity materials, such as ice or plastic, these results indicate that convective transport in the fluid is more efficient than conduction through the solid. The CHT simulations augment the experimental findings by enabling the calculation of $St$ accounting for spatial variations in heat flux that cannot be measured directly in experiments. These results underscore the importance of accounting for heat conduction when analyzing convective heat transfer over low-conductivity rough surfaces. Comparisons with isothermal wall simulations further reveal that neglecting solid conduction leads to an overprediction of $St$, particularly at the roughness crests.

Given the low thermal conductivity of ice, these conclusions suggest that incorporating heat conduction into the analysis is important for accurate heat transfer prediction. Moreover, the presence of roughness on iced surfaces implies that spatial variations in temperature and heat flux could significantly influence the ice accretion process. However, it is important to recognize differences between the present simulations and real-world icing scenarios. In the current setup, heat is supplied from the bottom of the plate, whereas on iced airfoils, heat originates at the ice surface due to the release of latent heat during solidification. Additionally, roughness not only modifies local heat transfer but also influences the impingement of supercooled droplets and the dynamics of water runback, if present. These distinctions may affect the role of conduction in the overall heat transfer process and warrant further investigation. Nonetheless, the present simulations validate the CHT/WMLES methodology and establish a foundation for future studies aimed at characterizing heat transfer on iced airfoils under realistic icing conditions.

\begin{acknowledgments}
This investigation was funded by the Boeing Co. and the NASA's TTT Program.
This research used resources of the Oak Ridge Leadership Computing Facility, which is a DOE Office of Science User Facility. The authors would like to thank Stephen T. McClain from Baylor University for providing the experimental data and for valuable discussions. We also thank Salvador Gomez and Shilpa Vijay for their insightful comments on this article. 
\end{acknowledgments}

\appendix

\section{Evolution of solid temperature with thermal inertia}\label{Ap:betaEffect}

\begin{figure}
    \centering    
    \includegraphics{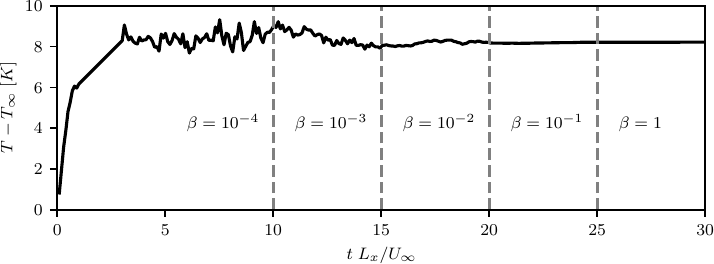}
    \caption{Evolution of surface temperature at $x/L_x=z/L_z = 0.5$ throughout the simulation for varying thermal inertia.}
    \label{fig:temp_evolution}
\end{figure}

Figure~\ref{fig:temp_evolution} shows the evolution of surface temperature at the center of the plate \((x/L_x = z/L_z = 0.5)\) for different values of \(\beta\). At the start of the simulation, the surface temperature rises rapidly due to the artificially low thermal inertia of the solid, set to \(\beta=10^{-4}\) times its real value. A statistically stationary state is reached at \(L_x/U_\infty \approx 3\). However, significant surface temperature fluctuations are observed for low \(\beta\) values. These fluctuations arise because the solid, with low thermal inertia, responds to temperature variations in the fluid. As \(\beta\) increases, the magnitude of these fluctuations decreases, eventually stabilizing when the solid’s thermal inertia matches its real value (\(\beta = 1\)).

\section{Spanwise averaged velocity profiles}\label{Ap:velProfiles}

\begin{figure}
    \centering    
    \includegraphics[width=\linewidth]{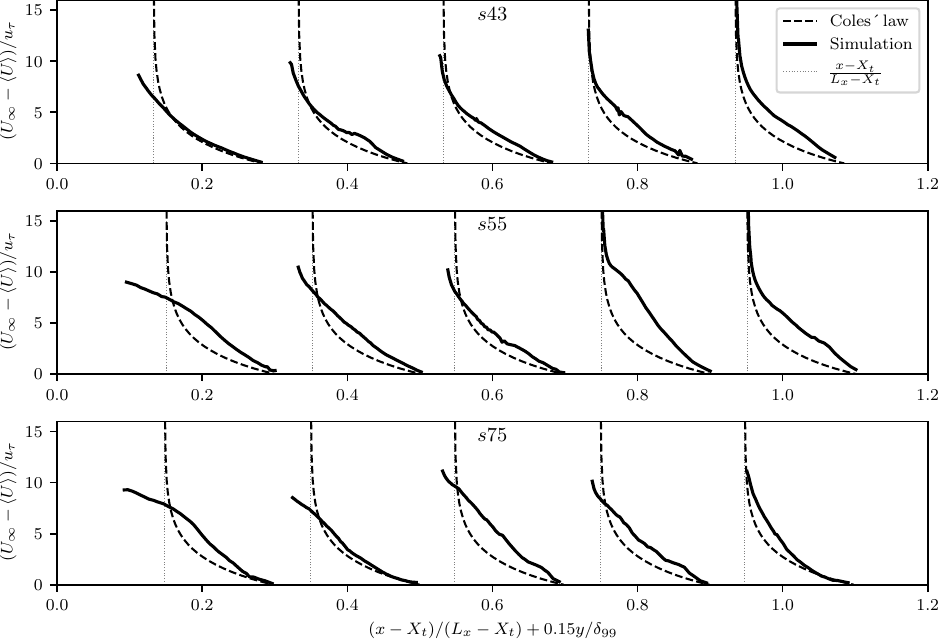}
    \caption{Velocity deficit profiles for the three geometries at different streamwise locations.}
    \label{fig:vel-def_profiles}
\end{figure}

Figure~\ref{fig:vel-def_profiles} presents spanwise-averaged velocity deficit profiles for cases $s43$, $s55$ and $s75$, compared to their smooth wall counterpart. The velocity profiles for all three roughness geometries fail to collapse onto the smooth wall profile, indicating a clear breakdown of outer-layer similarity. This deviation persists even in the downstream regions of the plate, where $k_z/\delta_{99}$ has diminished considerably. These results demonstrate that the influence of large roughness elements extends throughout the entire computational domain, affecting the flow field up to the boundary layer edge. This behavior is consistent with the observations reported by Hu et al.~\cite{xiaohanhuFlowIceRoughness2024} for flow over a discrete roughness patch, where the effects of roughness extend a significant distance downstream of the end of the roughness patch.

\bibliography{library_bibtex}

\end{document}